\documentclass[journal,twoside,print]{ieeecolor}
\usepackage{generic}
\usepackage{cite}
\usepackage{amsmath,amssymb,amsfonts}
\usepackage{algorithmic}
\usepackage{graphicx}
\usepackage{algorithm,algorithmic}
\usepackage{hyperref}
\hypersetup{
    colorlinks=true,
    linkcolor=black,
    citecolor=green,
    urlcolor=cyan,
    hidelinks=false
}
\usepackage{textcomp}
\usepackage[nameinlink,capitalize]{cleveref}
\usepackage{subfigure}

\usepackage{xcolor}
\usepackage{theorem}

\usepackage{comment}
\usepackage{booktabs}

\usepackage{cancel}

\newcommand{\CC}{\ensuremath{\mathbb{C}}}
\newcommand{\RR}{\ensuremath{\mathbb{R}}}

\newcommand{\imag}{{\imath}}

\definecolor{SeabornBlue}{HTML}{1f77b4}
\definecolor{SeabornOrange}{HTML}{ff7f0e}
\definecolor{SeabornRed}{HTML}{d62728}
\definecolor{SeabornGreen}{HTML}{2ca02c}

\def\BibTeX{{\rm B\kern-.05em{\sc i\kern-.025em b}\kern-.08em
    T\kern-.1667em\lower.7ex\hbox{E}\kern-.125emX}}
\begin{document}

\title{A Complex UNet Approach for Non-Invasive Fetal ECG Extraction Using Single-Channel Dry Textile Electrodes}

\author{Iulia Orvas, Andrei Radu, Alessandra Galli, Ana Neacșu, Elisabetta Peri 
\thanks{ Iulia Orvas and Ana Neacșu are with the SIGMA Laboratory, CAMPUS Institute, National University of Science and Technology Politehnica
Bucharest, Bucharest, Romania.}
\thanks{Andrei Radu is with the SIGMA Laboratory, CAMPUS Institute, National University of Science and Technology Politehnica
Bucharest, Bucharest, Romania 
and also with the Department of Information Engineering and Computer Science, University of Trento, Trento, Italy. Corresponding author: Andrei Radu (email: andrei\_radu.danila@upb.ro; andreiradu.danila@unitn.it).}
\thanks{Alessandra Galli and Elisabetta Peri are with the Department of Electrical Engineering, Eindhoven University of Technology, Eindhoven, The Netherlands.}
\thanks{Alessandra Galli is supported by the European Union’s Horizon Europe research and innovation programme under the Marie Skłodowska-Curie Postdoctoral Fellowship, project no. 101063008.}}

\maketitle

\begin{abstract}
\textit{Objective}: Continuous, non-invasive pregnancy monitoring is crucial for minimising potential complications. The fetal electrocardiogram (fECG) represents a promising tool for assessing fetal health beyond clinical environments. Home-based monitoring necessitates the use of a minimal number of comfortable and durable electrodes, such as dry textile electrodes. However, this setup presents many challenges, including increased noise and motion artefacts, which complicate the accurate extraction of fECG signals. To overcome these challenges, we introduce a pioneering method for extracting fECG from single-channel recordings obtained using dry textile electrodes using AI techniques.
\textit{Methods}: We created a new dataset by simulating abdominal recordings, including noise closely resembling real-world characteristics of \textit{in-vivo} recordings through dry textile electrodes, alongside maternal electrocardiogram (mECG) and fECG. To ensure the reliability of the extracted fECG, we propose an innovative pipeline based on a complex-valued denoising network, Complex UNet ($\mathbb{C}$UNet). Unlike previous approaches that focused solely on signal magnitude, our method processes both real and imaginary components of the spectrogram, addressing phase information and preventing incongruous predictions.
We evaluated our novel pipeline against
traditional, well-established approaches, on both simulated (\textit{in-silico}) and real (\textit{in-vivo}) data in terms of fECG extraction and R-peak detection.
\textit{Results}: The results showcase that our suggested method achieves new state-of-the-art results, enabling an accurate extraction of fECG morphology across all evaluated settings.
\textit{Significance}: This method is the first to effectively extract fECG signals from single-channel recordings using dry textile electrodes, making a significant advancement towards a fully non-invasive and self-administered fECG extraction solution.
\end{abstract}


\section{Introduction}
\label{sec:introduction}
Intrauterine death (stillbirth) represents one of the major public health issues, with an estimated global incidence of approximately 2 million cases per year~\cite{hug2020neglected}. Pregnancy complications, which can lead to adverse outcomes such as stillbirth, affect around 20\% of pregnancies \cite{valderrama2020review, ananth2006maternal}.
Many of these negative outcomes could be prevented through extensive and continuous pregnancy monitoring. Consistent and long-term monitoring could allow for the early detection of physiological abnormalities, facilitating timely intervention and thus reducing risks for both the mother and the fetus. In particular, home-based monitoring could offer significant benefits by reducing the frequency of hospital visits while ensuring continuous surveillance.

Currently, the gold standard for fetal monitoring is cardiotocography (CTG), a method used to measure fetal heart rate (fHR) and uterine contractions. However, CTG has several limitations, including requiring specialised medical staff to position the probe and its unsuitability for long-term use \cite{alfirevic2017continuous, hamelmann2019fetal}.
These constraints make CTG inadequate for continuous and home-based monitoring.

In recent years, non-invasive fetal electrocardiographic (fECG) monitoring has been proposed as a promising alternative to CTG~\cite{clifford2011clinically}. fECG can be acquired through electrodes placed on the maternal abdomen, providing a more detailed signal. Indeed, unlike CTG, fECG records the full electrocardiographic trace, offering crucial information not only about the fHR but also about its variability and morphology, both of which are key indicators of fetal well-being. For instance, fECG can detect abnormalities such as a shortened QT interval, which is associated with fetal hypoxia~\cite{kahankova2019review}, or measure the duration of the QRS complex, which can indicate intrauterine growth restriction (IUGR) by reflecting changes in fetal heart size~\cite{smith2018systematic}.

However, extracting fECG from maternal abdominal signals presents considerable technical challenges. 
The presence of unwanted artefacts in the signal often results from the acquisition process, where surface electrodes capture more than the signal of interest. The main interferences come from the difference in power of the fECG, compared to the maternal counterpart, muscle motion activity (commonly from electromyogram or electrohysterogram), or power grid-specific noise.
Effectively removing these disturbances is critical to obtaining a clean, interpretable fECG signal. The most challenging step is the removal of the maternal ECG (mECG), as its frequency content significantly overlaps with that of the fECG, making traditional filtering techniques inadequate. Over the years, various denoising techniques have been developed to extract a reliable fECG signal, including filtering, template subtraction, and signal decomposition. Adaptive filtering techniques include methods like Kalman filtering~\cite{de2024unsupervised} and more advanced nonlinear filters, such as those leveraging machine learning or deep learning~\cite{sharma2021fecg,wahbah2024deep}. In \cite{Galli2021}, the authors apply a technique called template subtraction, in which an approximation of the mECG is constructed based on the QRS complexes in the recording and subtracted from the original signal in order to obtain the fECG. From the signal processing domain, more specifically, from signal decomposition, multiple methods for Blind Source Separation (BSS) can be adapted for this task. Examples of these classical methods are presented in \cite{galli2022automatic}, which use principal and independent component analysis (PCA, ICA). More advanced methods, such as periodic ones ($\pi$CA), have been proposed in works such as \cite{galli2024improved}.
Recently, deep-learning methods have been employed to solve the ill-posed inverse problem of fECG extraction with promising results. Zhong \textit{et. al} \cite{zhong2018deep} were pioneers in employing a convolutional neural network (CNN)-based model to directly detect the fetal R-peak using single-channel non-invasive fECG. Building on this foundation, Lee \textit{et al.} \cite{lee2018fetal} introduced a modified CNN that enhanced the original framework through the incorporation of a post-processing scheme and a deeper architecture, which enabled more comprehensive feature extraction. Lately, more complex UNet-like structures have been showing promising results in the medical segmentation tasks, and they have been adapted for R-peak detection \cite{zhou2024using}. In \cite{Huang2025tcgan} a novel GAN-based architecture (TCGAN) is proposed to effectively extract fetal ECG from abdominal recordings, enabling accurate fetal monitoring by preserving waveform details. 

Although numerous fECG extraction methods have been developed and fECG analysis holds significant potential, its use remains largely confined to clinical settings, with limited adoption for long-term, home-based monitoring. This is due to two major limitations. Firstly, current devices rely on large electrophysiological patches that cover a significant portion of the maternal abdomen, restricting the mother's movement and making long-term monitoring uncomfortable. Secondly, such devices typically employ traditional adhesive wet electrodes, like Ag/AgCl electrodes, which are not ideal for continuous monitoring because the gel can dry out over time, leading to signal degradation. Additionally, prolonged use of these electrodes may cause skin irritation, further limiting their suitability for prolonged monitoring. To address the first issue, recent developments have focused on using single-electrode systems, which reduce the bulk of the device. For example, Zhang and Wei~\cite{zhang2020complete} proposed an iterative, adaptive method utilising empirical mode decomposition (EMD) to decompose a single-channel signal into multiple components, effectively isolating the fECG signal. Similarly, Niknazar et al.~\cite{niknazar2012fetal} introduced an fECG extraction technique based on an extended nonlinear Bayesian filtering framework, which models all the sources contributing to the single-channel signal. In contrast, Aldamani et al.~\cite{almadani2023one} employed a different approach, utilising two parallel UNets with transformer encoding to separate mECG from fECG in single-channel recordings.
To tackle the limitations of wet electrodes, a promising solution lies in the development of dry or capacitive electrodes, made from advanced textiles or other innovative materials that improve comfort and minimise adverse effects. However, using these electrodes introduces new challenges, such as increased noise and motion artefacts \cite{Galli2021}, which make accurate fECG extraction even more difficult. As a result, current signal extraction techniques are not yet reliable enough for this type of acquisition.

In this context, our work seeks to fill this gap by proposing the first method for extracting fECG from single-channel recordings acquired using dry textile electrodes. As such, the main contributions of our work can be summarised as follows:
\textit{(i)} We created a new \textit{in-silico} dataset composed of fECG and mECG, with different levels of noise that appear in real-life scenarios when dry textile electrodes are used, to tackle the lack of availability of a public dataset suitable for deep neural network training, including fECG data recorded through dry electrodes. 
\textit{(ii)} We trained several state-of-the-art deep learning-based methods for extracting the fECG signal from the aforementioned mixture, enhancing the information provided by single-channel signals;
\textit{(iii)} We propose a new deep-learning architecture for fECG extraction, which better follows the morphology of these signals, as demonstrated by our results in both in-vivo and in-silico scenarios \footnote{\scriptsize An implementation will be made available upon acceptance.}. 

The paper is organised as follows: \cref{sec:methods} presents the methods used for the dataset creation as well as the architectures used for the neural networks. Extensive experiments and their results are reported in \cref{sec:results}, with \cref{sec:disc} dedicated to further explaining the performances of our proposed system. Finally, \cref{sec:conclusion} reports final remarks.

\IEEEPARstart{}{} 
\section{Materials and Methods}
\label{sec:methods}

\subsection{Datasets}
\label{sec:data}

\subsubsection{\textit{In-silico} Dataset}
\label{sec:silico}
This section details the key processing steps required to generate an \textit{in-silico} dataset. The procedure involves creating a dataset formed by the combination of synthetic fECG and mECG signals, deliberately corrupted with noise characteristic of dry-electrode recordings. The main steps include: i) extracting noise from real recordings collected from non-pregnant women, ii) characterising the extracted noise to develop a model that accurately represents its characteristics, and iii) synthesising the \textit{in-silico} dataset by adding the noise to synthetic fECG and mECG traces.

 \paragraph{Noise extraction} 9 recordings of non-pregnant women were included in the study. Each recording was composed of 4 dry textile electrodes positioned on the abdomen around the umbilicus. Each recording lasted 180 seconds.

In contrast to signals acquired with traditional wet electrodes, signals acquired through dry electrodes are contaminated by more sources of noise, including triboelectricity. This phenomenon is induced by charge transfer due to friction between the electrode and the skin, which is prevented by the presence of gel in wet electrodes. To reduce the common-mode noise, bipolar derivations were considered.

Obtaining the fECG is challenging due to noise and artefact components that overlap with its physiological frequency band (0–100 Hz). Therefore, we opted not to model noise outside this band and powerline interference, as these components are typically removed during the pre-processing stage. In the considered recordings, these components have been removed by applying a $10^{th}$ order Infinite Impulse Response (IIR) notch filter with a bandwidth of 1 Hz and a central frequency of 50 Hz. A $10^{th}$ order IIR bandpass filter is applied to focus on the frequency range of interest (1-100 Hz) that corresponds to physiological signals.  

The considered signals were recorded from non-pregnant women and thus consist solely of adult ECG and noise.  To isolate the noise component, the ECG signal must be removed from each channel separately. This step is conducted by using an algorithm previously proposed by Galli et al. \cite{Galli2021}. The algorithm identifies QRS complexes and removes the ECG by using BSS techniques. The remaining signal is composed of noise and artefacts only.

\paragraph{Noise characteristics} The characteristics of noise and artefacts have been investigated in the frequency domain and for each channel separately. 
Through simple visual inspections, we observed that the frequency content of the noise can be divided mainly into two parts: pink noise in the frequency range of 0-10 Hz, accompanied by white noise in the 10-100 Hz band. 

While these frequency bands were deduced to represent the main components of the noise, randomisation was introduced when simulating it to increase variability. Each type of noise was modelled separately to capture the variability observed across the recordings. For pink noise, the band was extended to range from 0 Hz to a random value between 9 and 12 Hz. White noise was simulated to a random value between 60 and 90 Hz.

The Gaussian mixture noise was modelled as a combination of two different types of Gaussian noise: background noise with a standard deviation of \(\sigma_0\) = 1 and impulse noise with a standard deviation of \(\sigma_1\) = 10. The mixture is controlled by the probability \(p\) = 0.1 of selecting the impulse noise at any given point.
    \[
    x_i = (\sigma_0 \cdot (1 - u_i) + \sigma_1 \cdot u_i) \cdot z_i
    \]
    where \( z_i \sim \mathcal{N}(0, 1) \) is a vector of standard normal random variables. This equation ensures that \( x_i \) has a standard deviation of \( \sigma_0 \) when \( u_i = 0 \) and \( \sigma_1 \) when \( u_i = 1 \).

Amplitude differences were also investigated. The noise spectrum amplitude for each type of noise was normalised and scaled during combination. The weights were selected based on empirical observations: pink noise was given the highest contribution (weight of 2), followed by white noise (weight of 0.2), and the Gaussian mixture noise (weight of 0.15). To simulate noise in the time domain, the following approach was used: the Fourier transforms of pink, white, and Gaussian-mixture noise were normalized separately, the three components were scaled by their respective weights and combined in the frequency domain, and finally, the inverse Fourier transform was then applied to obtain the noise signal in the time domain.

\begin{figure}
    \centering
    \includegraphics[width=1\linewidth]{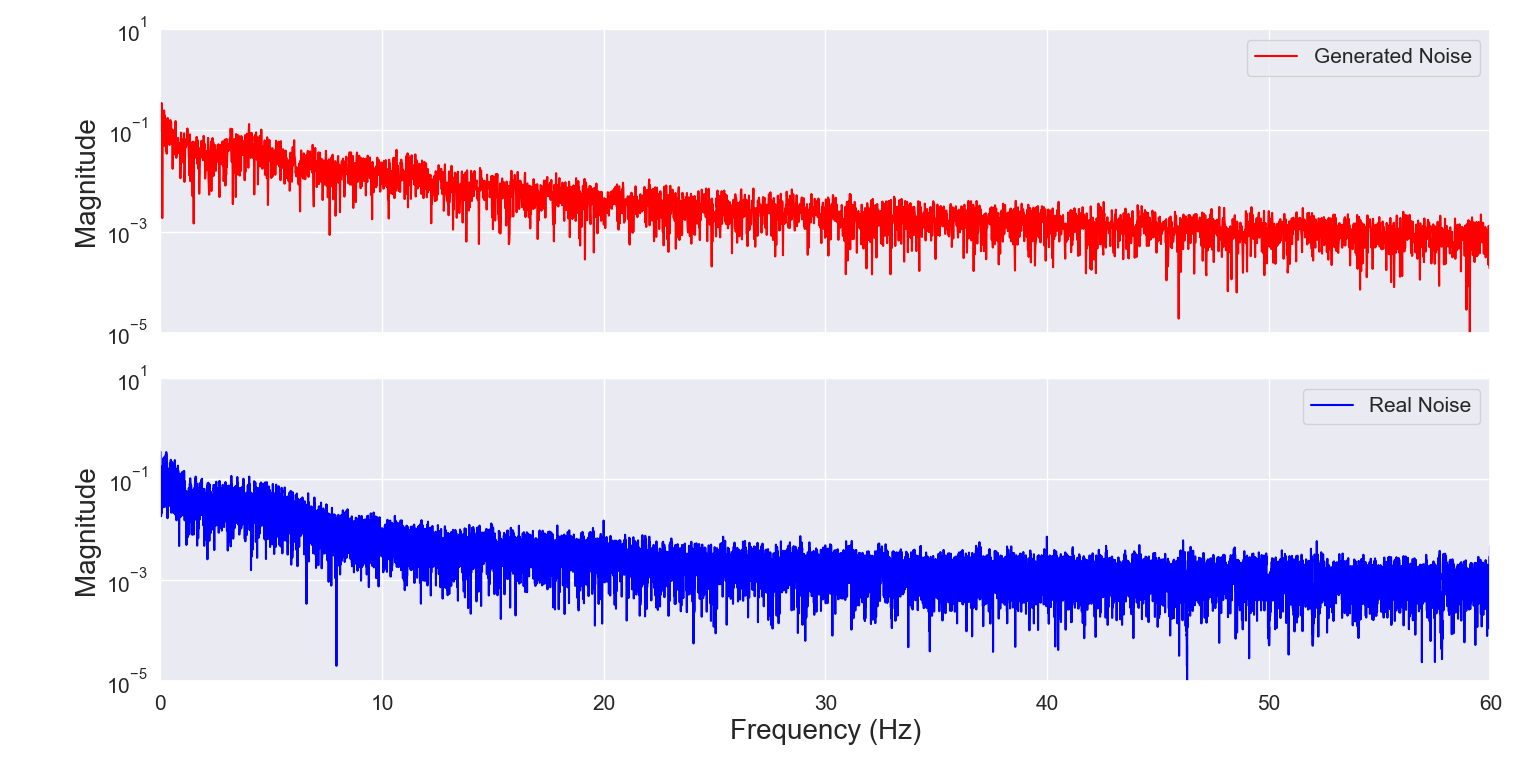}
        \caption{Illustrative case: generated noise signal in frequency domain}
    \label{fig:noise}
     \vspace{-0.6cm}
\end{figure}

In Figure \ref{fig:noise}, we report the power spectrum of an illustrative simulated noise signal, as well as an extracted noise signal, intended for replication.

\paragraph{Dataset generation} the synthetic dataset was obtained by superimposing 10.100 clean one-minute synthetic mECG and fECG signals as produced through \textit{fecgsyn} generator ~\cite{Andreotti2016} together with additive model-based noise, independently generated for each recording. 
For each acquisition, we select a random value from specific intervals, to mimic multiple gestational periods. Overall, as showcased in \cite{matonia2020fetal}, we set a maximum level amplitude of 20 $\mu$V.
The noise was considered additive, with SNR ranging from 5 to 20 dB, as derived from values reported in the literature~\cite{galli2024improved}, in order to replicate a realistic scenario. 

\begin{figure*}[htb]
    \centering
    \includegraphics[width=0.85\linewidth]{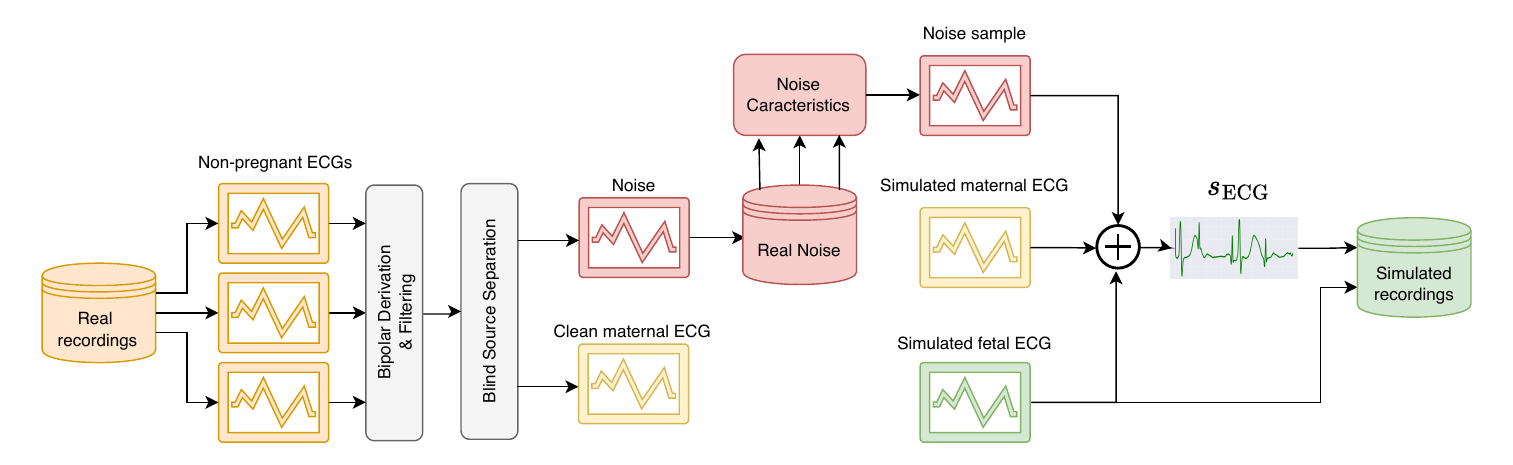}
    \caption{The dataset creation pipeline. We start by extracting the noise from real dry ECG recordings using signal processing and Blind Source Separation methods. After determining the noise characteristics, we sample noise alongside simulated maternal and fetal ECG to create our simulated dataset.}
    \label{fig:data-pipeline}
\end{figure*}

Part of this dataset, consisting of 100 simulated files, was reserved for testing and benchmarking the proposed method against existing approaches from the literature, while the remaining part was used to train the network.  Our pipeline for data generation is also presented in \cref{fig:data-pipeline}.

\subsubsection{In-vivo Datasets}
\label{sec:vivo}
In order to evaluate the performance of the proposed method in real-world scenarios, we used two additional datasets, one with wet electrodes and the second with dry electrodes.
\paragraph{PhysioNet dataset} The ``Noninvasive Fetal ECG—The PhysioNet Computing in Cardiology Challenge 2013'' dataset \cite{goldberger2000physiobank,silva2013noninvasive} consists of 75 one-minute abdominal recordings collected from multiple subjects using contact-wet electrodes. The recordings were obtained with various instrumentation setups, differing in frequency response, resolution, and configuration. For each acquisition, reference fetal QRS (fQRS) positions, derived from a fetal scalp electrode, are also provided. Despite not being acquired with dry electrodes, the PhysioNet dataset was included in this study as it serves as the standard benchmark for evaluating fECG extraction methods.
\paragraph{Dry Electrodes dataset} This dataset was acquired from four pregnant volunteers between 37 and 39 weeks of gestation. Each recording lasted between 21 and 24 minutes and was obtained using four dry textile electrodes. The study was approved by the Medical Research Ethics Committee of the Máxima Medical Centre, Veldhoven, the Netherlands (METC number: N19.061, 6 August 2019). Data collection took place at the hospital after obtaining signed informed consent from all volunteers.

\subsection{Model Architecture}
\label{sec:unet}
This section is dedicated to an overview of the proposed method. We first start by presenting the problem at hand, followed by a general description of the used architecture. Further sections are dedicated to a more in-depth analysis of the network's components.

\subsubsection{Problem formulation}
Deep Neural Networks (DNNs), increasingly prevalent across diverse fields like image classification, natural language processing, speech enhancement, etc., predominantly operate on real-valued inputs \cite{lee2022complex}. However, complex-valued neural networks (CVNNs) \cite{barrachina2023theory} offer significant advantages, particularly within the signal processing domain \cite{wu2023complex}, and are crucial for expanding the applications of DNNs in biomedical analysis \cite{weiss2022applications}. The ability to process complex-valued data is vital, as frequency analysis, a cornerstone of many signal processing techniques, inherently yields complex-valued outputs.  These complex values' magnitude and phase components carry distinct information, providing deeper insights into the signal's content. CVNNs demonstrate superior expressiveness compared to their real-valued counterparts \cite{barrachina2021complex}. For instance, end-to-end CVNNs have surpassed 2-channel real-valued networks in accelerated MRI reconstruction \cite{cole2020analysis}. In biomedical applications, incorporating phase information is crucial for advanced analyses and diagnostics. This is especially true for fECG extraction, where accurate timing and phase relationships are essential for meaningful interpretation and clinical insights \cite{mohammed2021survey}. 

\begin{figure*}[ht]
    \centering
    \includegraphics[width=\linewidth]{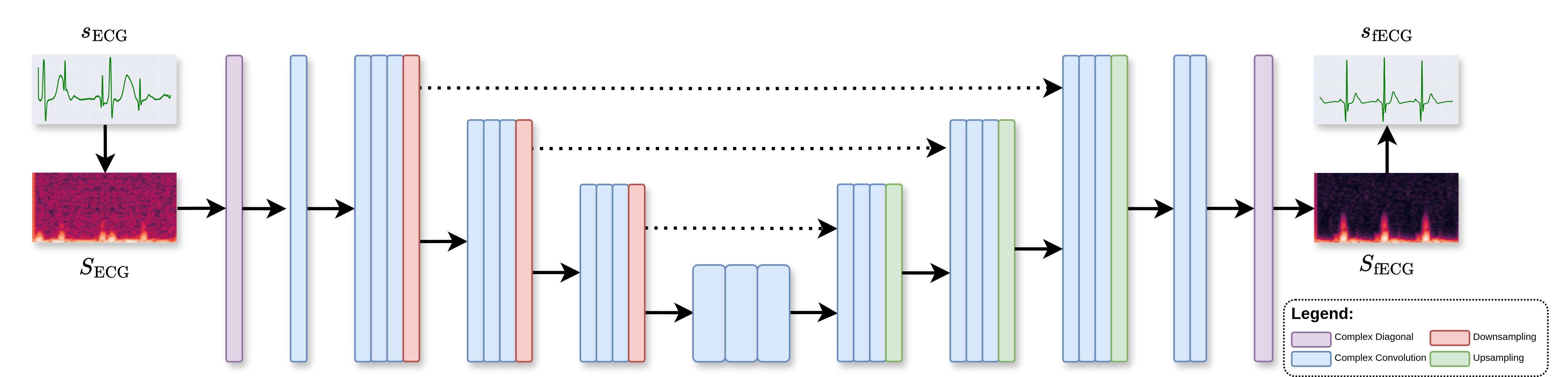}
    \caption{Our approach: First, we apply a time-frequency analysis on $s_\text{ECG}$ to obtain its spectrogram $S_\text{ECG}$. We propagate it through the ComplexUNet, denoted by $\mathcal{H}$, and obtain a denoised fetal extraction $S_\text{fECG} = \mathcal{H}(S_\text{ECG})$. The last step is composed of an inverse STFT, resulting in the fECG signal. 
    Note that each layer of our networks works in the Complex domain.}
    \label{fig:model-arch}
     \vspace{-0.5cm}
\end{figure*}

Incorporating phase information is crucial for advanced analyses and diagnostics for biomedical applications. This is especially true for fECG extraction, where accurate timing and phase relationships are essential for meaningful interpretation and clinical insights \cite{mohammed2021survey}. \cref{fig:model-arch} describes our system. We start by converting a single-channel abdominal signal ($s_\text{ECG} \in \RR^L$) into its spectrogram using short-time Fourier transform (STFT) $s_\text{ECG} \in \mathbb{R}^{(L)}$. Here, $L$ denotes the signal length, and $F$ and $T$ indicate the number of frequency and time bins, respectively. Our method is designed to effectively perform the necessary predictions on single-channel data, emphasising its suitability for fECG extraction.

\subsubsection{Complex-UNet}

UNet \cite{ronneberger2015unet}, a versatile architecture extensively applied in various computer vision tasks, served as a foundational inspiration for our proposed method.
Our work extends the application of UNet architectures to complex-valued inputs, opening new avenues for more accurate biomedical signal processing. By leveraging complex-valued data, our method capitalises on both magnitude and phase information inherent in the fECG signals, which is crucial when dealing with intricate and noisy biomedical environments. This adaptation allows for a more comprehensive capture of the signal's spectral properties, facilitating better isolation and extraction of fetal ECG components while effectively distinguishing them from maternal and noise-induced elements. 

We kept the overall architecture of UNet, adapting its components for our specific task. The UNet consists of three parts: the \textit{Encoder}, \textit{Bottleneck}, and \textit{Decoder}. The Encoder abstracts the input data into a compact latent space and builds a hierarchical representation. The Bottleneck processes these abstract features, transforming them into valuable information for the Decoder. Finally, the Decoder generates the segmentation mask by integrating high-level context from the Bottleneck with detailed features aggregated from the skip connections of the Encoder. 
 To leverage the full information contained in the complex form signal while using classic backpropagation and autodifferentiation techniques—given the challenges associated with autodifferentiation on complex-valued data—we adapted the standard layers of UNet to handle the real and imaginary parts separately. This separation ensures compatibility with classic training practices in neural networks.

We say $\CC$UNet is a complex-valued UNet with $m$ complex convolutional layers, considering for every $i \in \{1, \dots, m\}$ that there exists $N_i$, $N_{i-1}$, the number of input and output channels, respectively. Let $W_i \in \CC^{N_{i-1} \times N_i \times k \times k}$ be the associated complex weight tensor, parametrized by a kernel of size $k \neq 0$, the bias term $b_i \in \CC^{N_i}$, and the nonlinear activation operator $R_i: \CC^{N_i} \mapsto \CC^{N_i}$. If we denote by $W_i^{\mathbb{R}\rm{e}}$ and $W_i^{\mathbb{I}\rm{m}}$ the real and imaginary parts of $W_i$, each convolution layer maps the complex input as follows.
\begin{equation}
Y = R_i(W_i^{\mathbb{R}\rm{e}} \ast \mathbb{R}\mathrm{e} (\mathrm{x}) + \imag (W_i^{\mathbb{I}\mathrm{m}} \ast \mathbb{I}\mathrm{m} (\mathrm{x})))
\label{eq:complex-conv}
\end{equation}
 For a more straightforward approach, we group these complex convolution operators in $\CC$Down blocks and $\CC$Up blocks, for the Encoder and Decoder, respectively. Each block contains a sequence of complex convolutional layers and normalisation ones, ending with a downsampling or upsampling operator. Aligning our architecture with the original one, we kept a sampling factor of 2 for each block. 

Another important aspect is represented by the presence of skip connections between the Encoder layers and the associated Decoder layers. Introduced in \cite{he2015deepresidual}, skip connections, also known as residuals, have proved their efficiency, especially in extremely deep architectures, by allowing the gradients to apply significant changes even in the first parts of a neural network. UNet architectures use Encoder residuals by concatenating them with the abstract representation of the Decoder to preserve the fine-grained information present in the initial stages of feature engineering.

\subsubsection{Diagonal Layer}
Inspired by the network proposed in \cite{neacsu2022design}, to offer our network more expressivity, we introduced a special layer called the Diagonal layer in our architecture. This layer performs phase shifts in the complex plane, which are optimised during the training process. The weights of this layer consist of vectors $(\beta_{1,k})_{1\le k \le N_{i}}\in [0,2\pi[^{N_{i}}$, which are the main diagonal of the weight matrix. 

\begin{equation}
    \Lambda = \text{Diag}(e^{\imag \beta_{i,1}}, \dots, e^{\imag \beta_{i, F\times T}})
\label{eq:diag-w}
\end{equation}
In our experiments, we employed the diagonal layer as the first and last layer, \textit{i.e.} $i=\{1, m\}$.

Another important aspect of the proposed layer is the preservation of feature independence since all the extracted features are multiplied individually with the learned weights. This method is especially effective in downstream tasks on large pre-trained models \cite{wu2023connecting}.

\subsubsection{Complex Activation Functions}

 There are primarily two methodologies for developing complex activation functions. The first approach involves the use of split-complex operators, which represent an extension of classical real-valued operators and are applied individually to the real and imaginary components. An illustrative example belonging to this category is the Complex ReLU activation function, defined as follows:

\begin{equation}
 (\forall \zeta \in \CC)\quad    \mathbb{C}\text{ReLU}(\zeta) = \text{ReLU}(\mathbb{R}\text{e} \zeta) + \text{ReLU}(\mathbb{I}\text{m} \zeta)
    \label{eq:complex-relu}
\end{equation}

This results in decoupling the real and imaginary parts, which can lead to magnitude and phase distortions, breaking complex relationships but allowing for a better interpretability of the two. However, such concern can be attributed to applying such an activation function as a stand-alone operator. In the context of machine learning, the training process ensures that this relationship is preserved, per its importance. 

The second strategy is to use activation functions that operate jointly on the real and imaginary parts of their input. 

\textbf{Georgiou-Katsageras} \cite{georgiou1992complex} is a widely used function that preserves the structure of the complex domain, defined as:
\begin{equation}
    (\forall \zeta \in \CC)\quad    \text{GK}(\zeta) = \frac{\zeta}{ 1 + |\zeta|}.  
    \label{eq:gk-activation}
\end{equation}

\textbf{GroupSort} is another activation function that is not phase-preserving, presented in \cref{eq:gs-activation}. 

\begin{equation}
   (\forall \zeta \in \CC)\quad  \text{GS}(\zeta) = \min (\mathbb{R}\text{e} \zeta, \mathbb{I}\text{m} \zeta ) + \imag \max ( \mathbb{R}\text{e} \zeta, \mathbb{I}\text{m} \zeta ) 
    \label{eq:gs-activation} 
\end{equation}

While this function introduces phase and magnitude distortions, it has the advantage of encoding the meaningful real and imaginary components in higher representations, similar to max-pooling operations. Moreover, this activation, like Complex ReLU, treats the real and imaginary parts symmetrically, which is useful in our task since no part is inherently more important or contains more valuable information. 

We also performed convex combinations of the aforementioned activation functions to form new activation operators. For example, let us define the following complex function:

\begin{equation}
 (\forall \zeta\in \CC)\quad   \rho(\zeta) = 
        \mu_1  \mathbb{C}\text{ReLU}(\zeta) + 
        \mu_2 \text{GK}(\zeta) + 
        \mu_3 \text{GS}(\zeta)
    \label{eq:ro-activation}
\end{equation}
with  $\mu_1 + \mu_2 + \mu_3 = 1$ and $\mu_j \in [0, 1], \ \forall j \in \{1, 2, 3\}$ where $\mu_j$ represent learnable parameters during the training process.

Combining multiple activation functions allows for a multiple focus on different aspects of the input data.
To respect the desired complex-domain restrictions and functional properties of our activations, the combination of these must be crafted carefully. To this extent, we set the $\mu_j$ parameters as learnable by the network. As such, in each distinct layer, the optimisation process decides if the best approach is to focus on the decoupling of real and imaginary parts provided by Complex ReLU, the preservation of the phase offered by GK or the higher nonlinearity and symmetry of GS. This approach fosters a more adaptable behaviour, which is particularly crucial in our context, as it leverages the examination of both the magnitude and phase components while ensuring robust adaptation to the varying input characteristics of the analysed signal.

\subsection{Methods for comparison}
The presented solution was benchmarked against single-channel approaches identified in earlier studies, explicitly with the Extended Kalman Filter and Smoother (EKF and EKS) \cite{sarafan2022novel}, Singular Value Decomposition (SVD) for template subtraction, and Convolutional Neural Networks, such as UNet and AttUNet. The EKF is a nonlinear adaptation of the traditional Kalman Filter for systems governed by nonlinear dynamic equations. The nonlinear framework characterising the shape of the ECG signal was introduced by \cite{mcsharry2003dynamical}. The EKS operates similarly but applies the forward and backwards filters, yielding a refined estimate at the cost of method causality. In-depth information on these baseline algorithms is provided in \cite{uriguen2015eeg}.

In the template subtraction technique applied in \cite{varanini2014efficient}, the ECG signal is divided into individual cardiac cycles using the positions of the QRS complexes as reference points, and SVD is then employed to reconstruct a cleaner version of the ECG. This process is carried out twice: initially to eliminate the maternal ECG, and afterwards to reduce remaining noise in the initial fECG signal. Techniques such as filtering and template subtraction prove highly effective for denoising ECG signals recorded from a single channel \cite{suganthy2021detection}. 

Approaches based on BSS, such as ICA or $\pi$CA, were not considered, as they rely on the redundancy from simultaneous recordings across multiple electrodes and, therefore, cannot be applied to single-channel acquisitions.

 Recently, UNet architectures have been successfully employed in fetal ECG (fECG) analysis, particularly for fetal R-peak detection \cite{vijayarangan2020rpnet} and fECG extraction \cite{rahman2023fetal}. Furthermore, innovations like the Attention UNet (AttUNet), which incorporates attention mechanisms through attention gates, have advanced results in medical image segmentation tasks \cite{oktay2018attentionunet}.

\subsection{Evaluation Metrics}
\label{sec:metrics}
The proposed method was evaluated on two key tasks: R-peak detection, essential for determining fHR, and fECG signal extraction, which can be used to extract morphological parameters that provide insights into the overall health status of the fetus.

\subsubsection{R-peak detection}
The accuracy of R-peak detection was assessed by comparing the detected peaks with reference annotations obtained from simulated fECG for the \textit{in-silico} dataset and from the scalp fECG for the \textit{in-vivo} PhysioNet dataset. Each detection was labelled as a True Positive (TP) if the R-peak was detected, or a False Negative (FN) if it was missed. False Positives (FP) represented instances where a non-R-peak was incorrectly marked as an R-peak. An R-peak was deemed correctly detected if its estimated position was within 50 ms of the reference annotation, as outlined in~\cite{Association1994}.

Sensitivity and F-score metrics were used to assess the algorithm’s accuracy in detecting R-peaks. Sensitivity (SE) measures the proportion of actual R-peaks that are correctly identified: 

\begin{equation}
\text{SE} = \frac{\text{TP}}{ \text{TP}+\text{FN}}
\end{equation}

On the other hand, the F-score, which ranges from 0 to 1, reflects the overall detection performance, with a value of 1 indicating perfect detection accuracy:

\begin{equation}
\text{F-score} = \frac{2\cdot \text{TP}}{2\cdot \text{TP}+\text{FN}+\text{FP}}.
\end{equation}

R-peaks are used to calculate the fHR, a key parameter in clinical practice for assessing fetal well-being. To evaluate the accuracy of the fHR estimation, we introduced an additional metric that measures the difference between the true fetal HR (fHR) and the algorithm’s estimated HR ($\hat{\text{fHR}}$):

\begin{equation}
\text{HR}_{\text{err}} = \sqrt{\frac{\sum_{l=1}^{N_R}{(\hat{\text{fHR}}_l-\text{fHR}_l)^2}}{N_R}},
\end{equation}

where $l$ is the index of the $l$-th beat and $N_R$ is the number of heartbeats in the recording. $\text{HR}_{\text{err}}$ measurement unit is bpm. The smaller the $\text{HR}_{\text{err}}$, the more accurate the fHR estimation. 

\subsubsection{fECG extraction}

The quality of fECG signal extraction can be quantitatively evaluated only in simulations, as it requires comparison with a reference signal, which is unavailable for \textit{in-vivo} signals. For \textit{in-vivo} data, a qualitative assessment can be performed by visually inspecting the signals extracted by different methods.

For the \textit{in-silico} dataset, however, extraction performance can be evaluated by computing the Percent RMS Difference (PRD) metric, where RMS denotes the root mean square operation, and the Pearson Correlation Coefficient (PCC). The PRD quantifies the degree of dissimilarity between the target $\text{fECG}$, and the prediction, $\hat{\text{fECG}}$: 

\begin{equation}
PRD = 100 \cdot \sqrt{\frac{\sum{ \left( \text{fECG}(t) - \hat{\text{fECG}}(t) \right) ^2}}{\sum{  \text{fECG}(t) ^2}}}
\end{equation}

On the other hand, $\text{PCC}$ metric quantifies the strength and direction of the linear relationship between $\text{fECG}$ and $\hat{\text{fECG}}$. It is defined as follows:

\begin{equation}
\text{PCC} = 100 \cdot \frac{\sum [\text{fECG}(t) \cdot \hat{\text{fECG}}(t)]}{\sqrt{\sum \text{fECG}(t)^2} \cdot \sqrt{\sum\hat{\text{fECG}}(t)^2}}
\end{equation}

We showcase the correlation between the two signals by considering the possible range of values for this metric, $\text{PCC} \in [-1, 1]$. While a $\text{PCC}=0$ indicates no linear correlation, a value closer to $\text{PCC}=\pm1$ demonstrates a perfect positive (or negative - which is not desired) correlation amongst the evaluated signals.

\section{Results}
\label{sec:results}
\subsection{In-silico results}
\begin{figure}
    \centering
    \subfigure[]{\includegraphics[width=\linewidth]{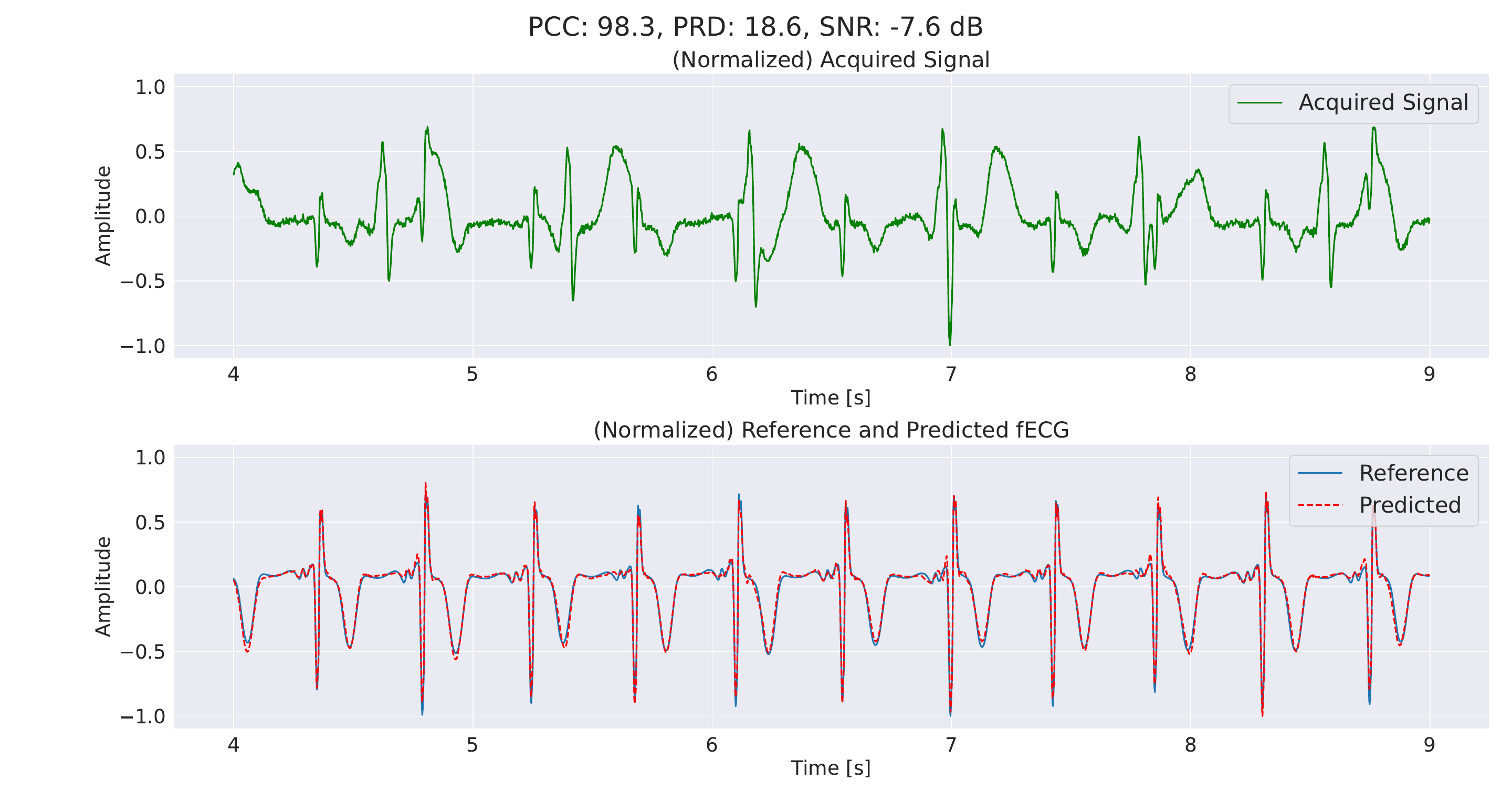}}
    \subfigure[]{\includegraphics[width=\linewidth]{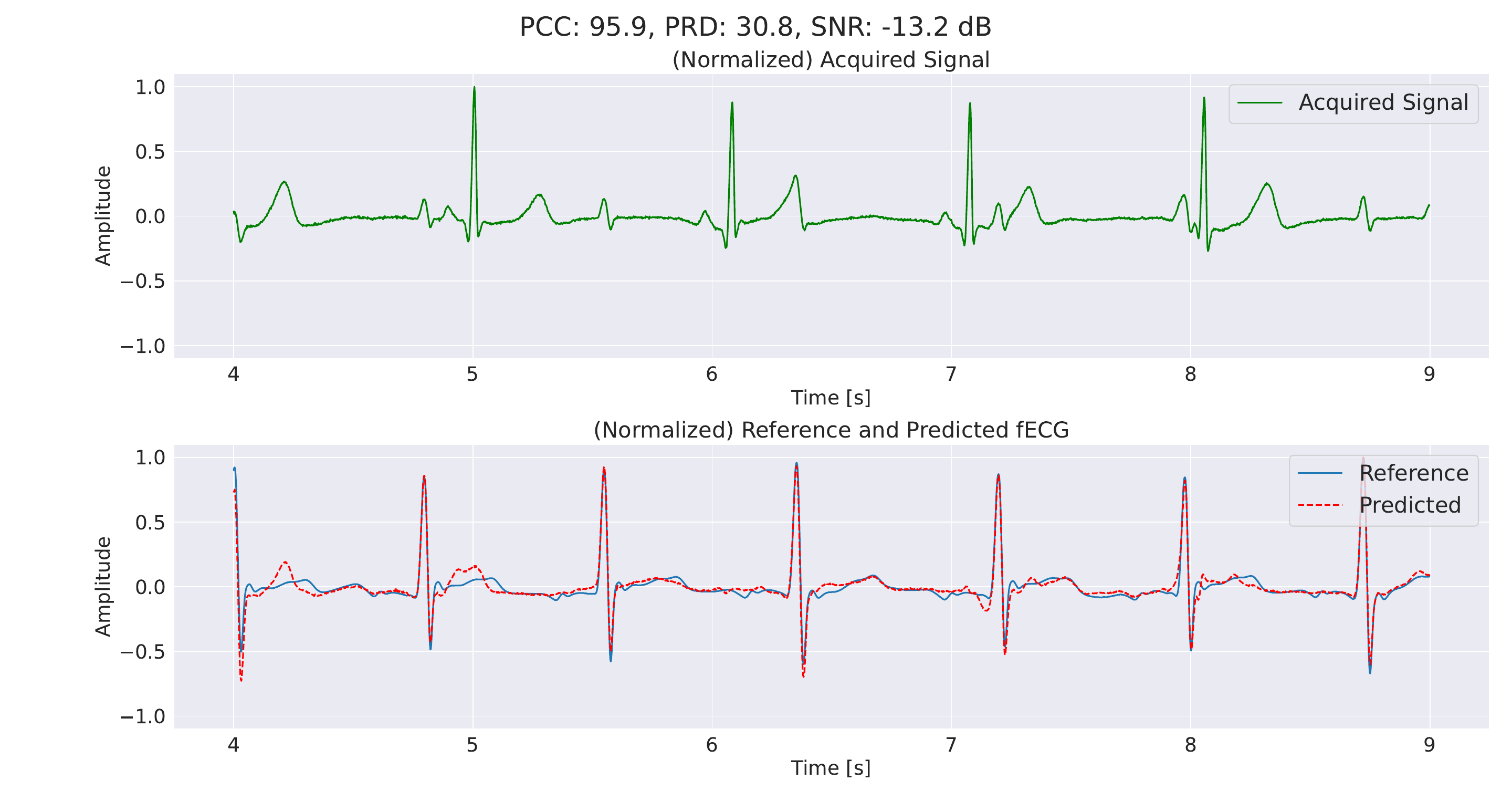}}
    \subfigure[]{\includegraphics[width=\linewidth]{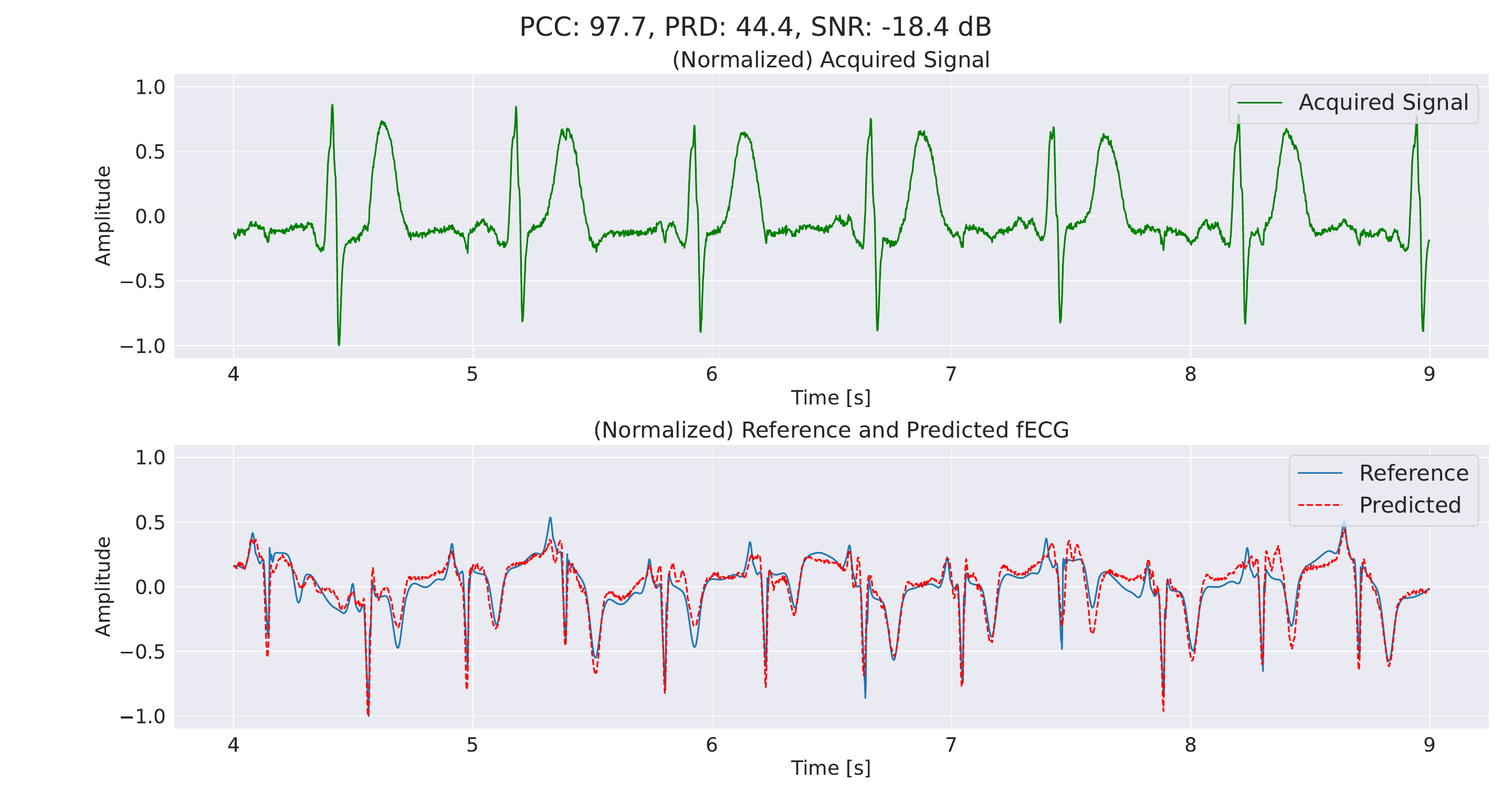}}
    \caption{Examples of abdominal simulated (noisy, in \textcolor{SeabornGreen}{green}) signals alongside the reference (\textcolor{SeabornBlue}{blue}) and predicted fECGs (\textcolor{SeabornRed}{red}), at different PRD levels: (a) low, (b) medium, (c) high. All signals are normalised for a better visualisation.}
     \label{fig:examples}
     \vspace{-0.6cm}
\end{figure}

\cref{fig:examples} shows the comparison between the noisy abdominal signal (top) alongside the predicted fECG signal and its reference (bottom), for different levels of noise. 

\begin{figure}
    \centering
    \includegraphics[width=\linewidth]{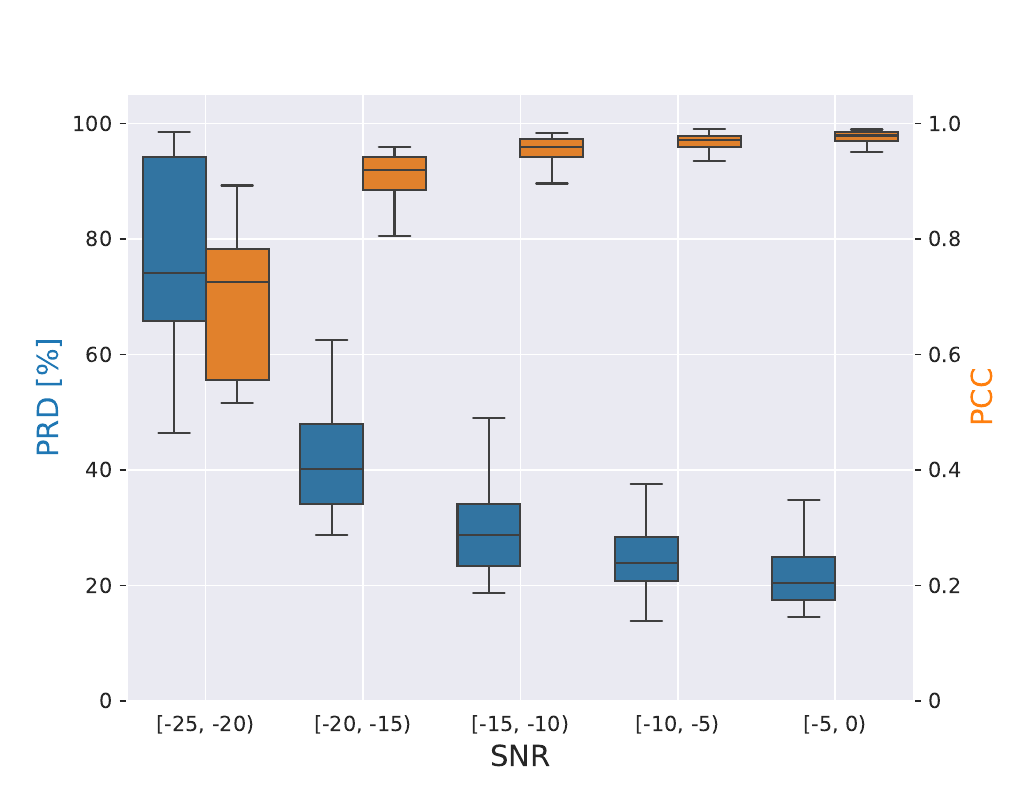}
    \caption{Boxplots of PRD (\textcolor{SeabornBlue}{blue}) and PCC (\textcolor{SeabornOrange}{orange}) values w.r.t. the SNR for \CC UNet}
    \label{fig:boxplot}
     \vspace{-0.5cm}
\end{figure}

\cref{tab:snr} and ~\cref{fig:boxplot} illustrate how the metrics vary with changes in SNR. In this analysis, SNR values range from $-25$ to $0$ dB, as only the fECG signal is considered a useful component, while the combined mECG, acquisition noise, and artefacts are treated as noise. As a result, high SNR values are obtained for fECG signals with greater amplitude, corresponding to advanced gestational ages, while low SNR values are characteristic of early pregnancy, where the fECG amplitude is reduced due to the small size of the fetal heart.  \cref{tab:snr} presents the mean and standard deviation of the PRD, PCC, F-score, and SE metrics, while \cref{fig:boxplot} depicts the distribution of the PRD and PCC metrics. As expected, increasing the SNR improves performance, resulting in a higher PCC and lower PRD.

\begin{figure*}
    \centering
    \includegraphics[width=\linewidth]{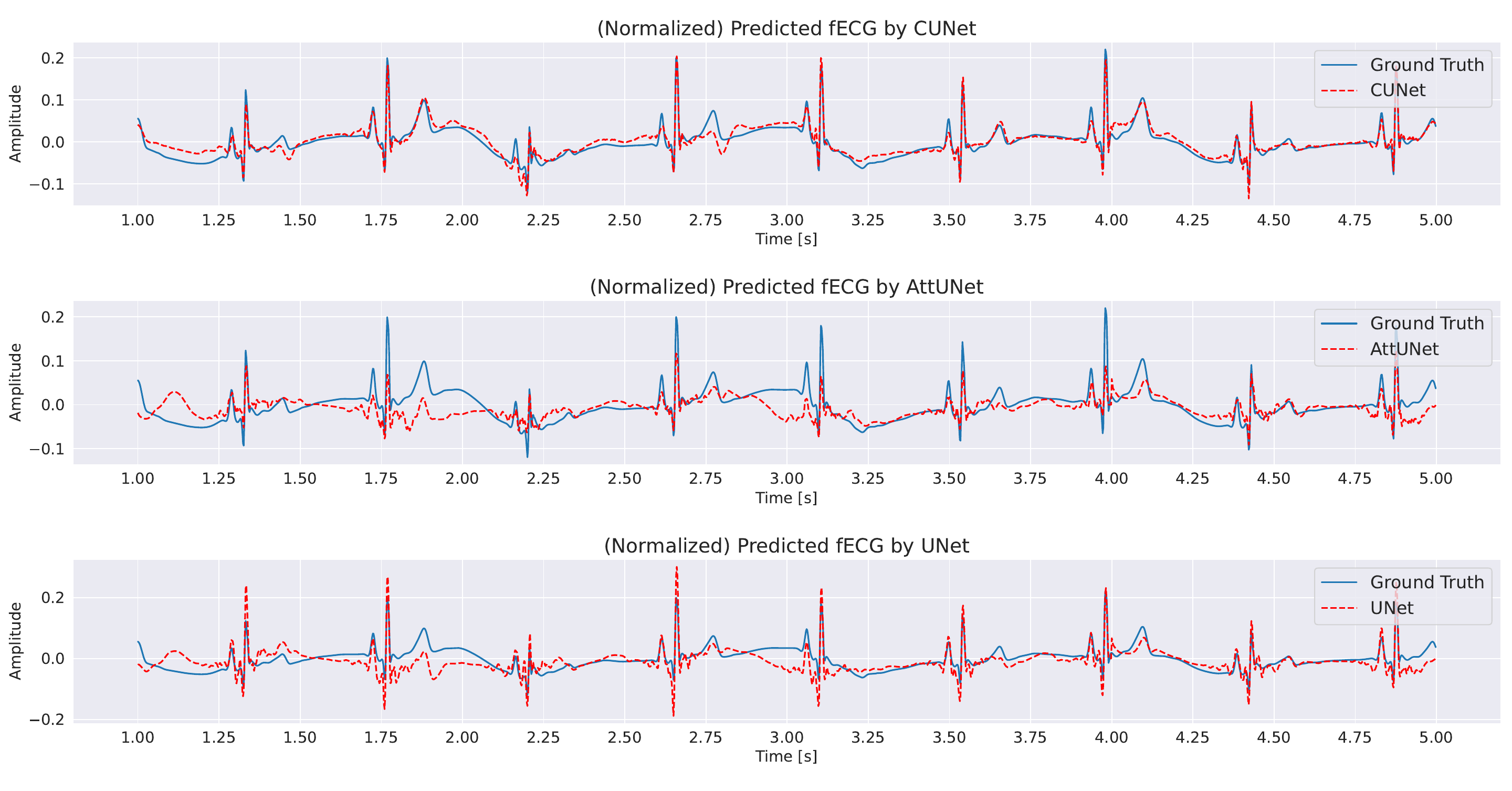}
    \caption{Visual comparison between the reference (\textcolor{SeabornBlue}{blue}) and predicted fECG (\textcolor{SeabornRed}{red}), using different neural-based methods. Top represents $\CC$UNet, middle is AttUNet, and bottom is UNet.}
    \label{fig:predicted-signals}
     \vspace{-0.5cm}
\end{figure*}

\cref{tab:comparison_silico} presents the performance metrics for fECG extraction (PRD and PCC) and R-peak detection (F-score, SE, and $\text{HR}_\text{err}$) across different methods, including traditional approaches (EKF, EKS, SVD) and deep learning-based models (UNet, AttUNet, and $\CC$UNet). The reported results were obtained on the entire \textit{in-silico} dataset without distinguishing between different SNR levels. The proposed $\CC$UNet method outperforms all other approaches across all evaluated metrics. It achieves the lowest PRD ($26.2 \pm 11.6$) and the highest PCC ($95.9 \pm 5.7$), indicating superior fECG signal reconstruction. Additionally, for R-peak detection, $\CC$UNet attains the highest F-score ($99.8 \pm 0.2$) and SE ($99.6 \pm 0.4$), demonstrating near-perfect detection capability. \cref{fig:predicted-signals} further illustrates a visual comparison between the reference and predicted fECG signals using different neural-based methods. The signals reconstructed by $\CC$UNet exhibit the closest alignment with the reference, preserving both amplitude and waveform morphology. In contrast, AttUNet and UNet show larger deviations, particularly around peak locations. This visual validation supports the quantitative results presented in \cref{tab:comparison_silico}.
 
\begin{table*}
\centering
\caption{ \scriptsize Average PRD values obtained by the proposed algorithm w.r.t. the SNR for $\CC$UNet}
\label{tab:snr}
\begin{tabular}{c|c|c|c|c|c}
\toprule
SNR &   {[}-25, -20) & {[}-20, -15) & {[}-15, -10) & {[}-10, -5) & {[}-5, 0) \\ \midrule
PRD $\downarrow$ & $80.39\pm24.63$ & $43.70\pm15.45$ & $30.75\pm9.50$ & $25.04\pm5.87$ & $21.89\pm5.06$ \\  \midrule
PCC $\uparrow$ & $64.65\pm22.93$ & $89.23\pm9.83$ & $94.97\pm3.80$ & $96.88\pm1.51$ & $97.80\pm1.13$ \\ \midrule
F-score $\uparrow$ &  $99.84\pm0.36$ & $99.81\pm0.21$ & $99.79\pm0.24$ & $99.79\pm0.24$ & $99.83\pm0.21$ \\ \midrule
SE $\uparrow$ & $99.69\pm0.71$ & $99.62\pm0.42$ &  $99.59\pm0.48$ & $99.58\pm0.48$ & $99.66\pm0.42$ \\
\bottomrule
\end{tabular}
\end{table*}

\begin{table*}
	\centering
	\caption{ \scriptsize Evaluation metrics for fECG extraction (PRD and PCC) and R-peak detection (F-score, SE, and $\text{HR}_{\text{err}}$) for the proposed method compared to benchmark algorithms. All values represent single-channel estimates computed on the simulated test set of 100 subjects. The best result is highlighted in bold for each metric, while the second-best is underlined. }
	\label{tab:metriche}
	\begin{tabular}{c|c|c|c|c|c}
		\toprule
		Method &\textbf{PRD} $\downarrow$  & \textbf{PCC} $\uparrow$ & \textbf{F-score} $\uparrow$ [\%] & \textbf{SE} $\uparrow$ [\%] & $\text{HR}_{\text{err}}$ $\downarrow$ [bpm]\\ 
		\midrule
		EKF & $79.6\pm45.2$ & $65.0\pm28.6$ & $88.7\pm27.0$ & $88.2\pm28.2$ & $8.0\pm17.5$\\
		EKS &  $79.0\pm45.1$ & $66.2\pm28.6$ & $88.7\pm28.0$ & $88.1\pm28.2$ & $8.1\pm17.7$\\
		SVD & $72.1\pm110.4$ & $75.9\pm31.6$ & $92.0\pm32.4$ & $90.2\pm28.3$ & $17.7\pm23.7$\\
        \midrule 
        UNet & $64.6\pm18.9$ & $\underline{77.4\pm13.4}$ & $\underline{99.3\pm1.6}$ & $\underline{98.8\pm2.6}$ & $1.24\pm1.47$\\
        AttUNet & $\underline{63.1\pm19.8}$ &  $77.4\pm15.0$&$98.4\pm 10.1$ & $98.2\pm9.0$ & $\underline{1.19\pm1.67}$\\
        \midrule
        $\CC$UNet & $\mathbf{26.2\pm11.6}$ & $\mathbf{95.9\pm5.7}$ & $\mathbf{99.8\pm0.2}$ & $\mathbf{99.6\pm0.4}$
        & $\mathbf{0.40\pm0.81}$ \\
		\bottomrule    
	\end{tabular}
    \label{tab:comparison_silico}
     \vspace{-0.5cm}
\end{table*}

\subsection{In-vivo results}
\subsubsection{PhysioNet dataset}

\begin{figure}[h]
	\centering
	\includegraphics[width=\linewidth]{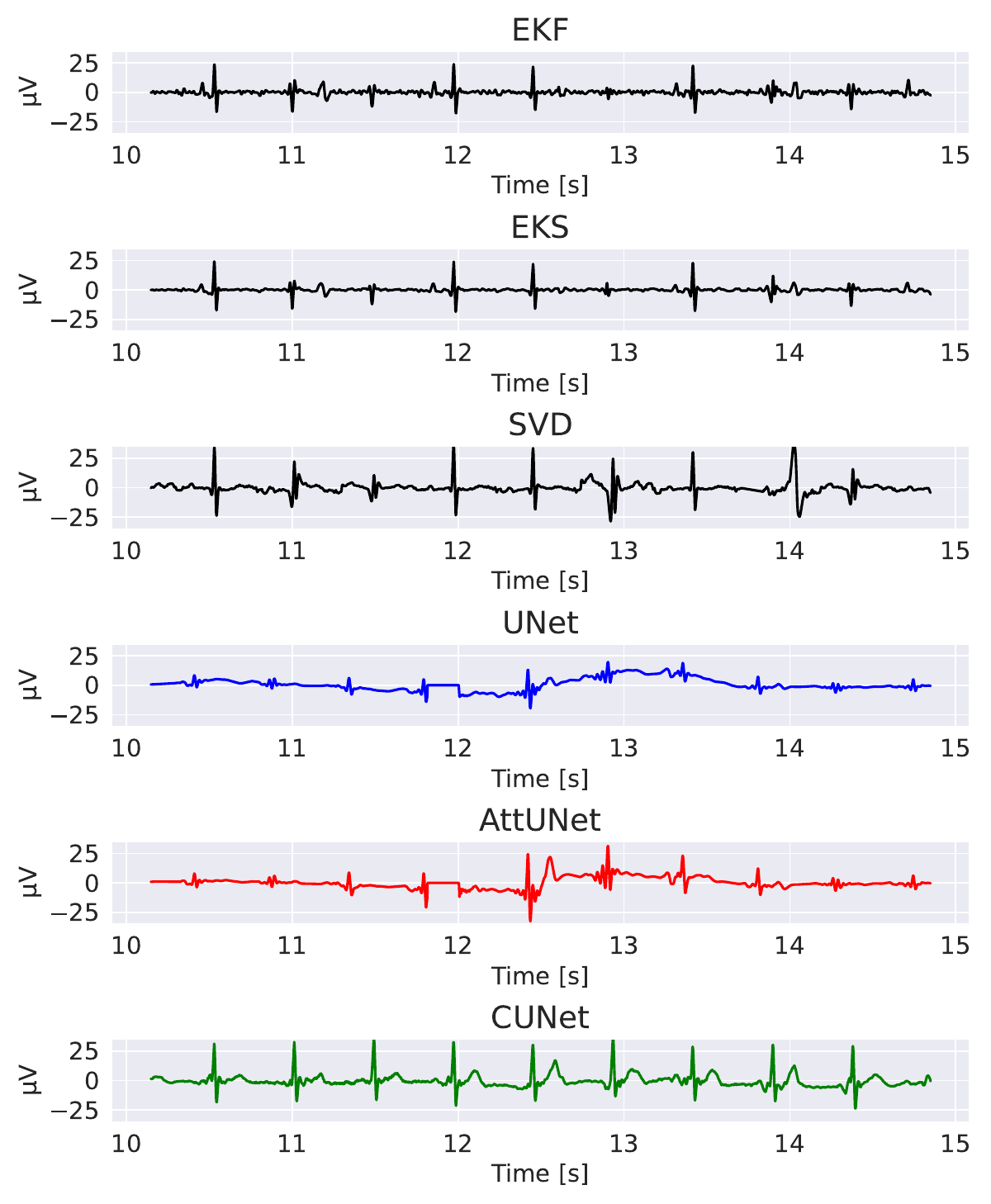}
	\caption{Visual inspection of the results obtained using UNet (\textcolor{SeabornBlue}{blue}), AttUNet (\textcolor{SeabornRed}{red}) and $\CC$UNet (\textcolor{SeabornGreen}{green}) against different benchmarking algorithms (black) on the PhysioNet dataset.}
	\label{fig:example_Physionet}
     \vspace{-0.5cm}
\end{figure}

\cref{tab:metriche-physionet} presents the evaluation metrics for R-peak detection, including F-score, SE, and HR for the proposed method compared to benchmark algorithms. The results are computed on the PhysioNet dataset using single-channel estimates. The proposed $\CC$UNet model achieves competitive performance, with an F-score of $77.8 \pm 18.6$ and SE of $72.1 \pm 22.5$, slightly lower than the best-performing model, EKF.  
Finally, \cref{fig:example_Physionet} provides an illustrative example of how different methods perform in extracting the fECG signal. The signals obtained with all the benchmark methods exhibit significant noise and reduced signal quality, making it difficult to distinguish the fECG components. Particularly, none of the estimations allow for the recognition of T-waves, and in the estimations provided by EKS, EKF, and UNet, even identifying the QRS complexes is challenging. In contrast, the proposed method (bottom panel) achieves superior signal quality, with a more clearly defined waveform and better visibility of the characteristic fECG features.  

\begin{table}
	\centering
	\caption{\scriptsize Evaluation metrics for R-peak detection (F-score, SE, and $\text{HR}_{\text{err}}$) for the proposed method compared to benchmark algorithms. All values represent single-channel estimates computed on the PhysioNet dataset. The best result is highlighted in bold for each metric, while the second-best is underlined.}
	\label{tab:metriche-physionet}
	\begin{tabular}{c|c|c|c}
		\toprule
		Method & \textbf{F-score} $\uparrow$ [\%] & \textbf{SE} $\uparrow$ [\%] & $\text{HR}_{\text{err}} \downarrow$ [bpm]\\ 
		\midrule
		EKF &  \textbf{80.0} $\pm$ \textbf{20.0} & \textbf{75.5} $\pm$ \textbf{23.8} & $15.4\pm18.6$\\
		EKS &  $77.8\pm20.9$ & \underline{$75.2\pm23.8$} & $15.3\pm18.3$\\
		SVD &  $75.6\pm26.3$ & $74.5\pm23.4$ & \textbf{13.6} $\pm$ \textbf{16.4}\\
        \midrule
        UNet & $54.3\pm21.0$ & $43.4\pm17.4$ & $23.2\pm18.4$\\
        AttUNet &  $51.6\pm22.4$ & $41.0\pm18.3$ & $22.0\pm18.2$\\
        \midrule
		$\CC$UNet  & \underline{$77.8\pm18.6$} & $72.1\pm22.5$ & \underline{$14.9\pm17.8$} \\
		\bottomrule   
	\end{tabular}
     \vspace{-0.5cm}
\end{table}

\begin{figure}
    \centering
    \subfigure[]{\includegraphics[width=\linewidth]{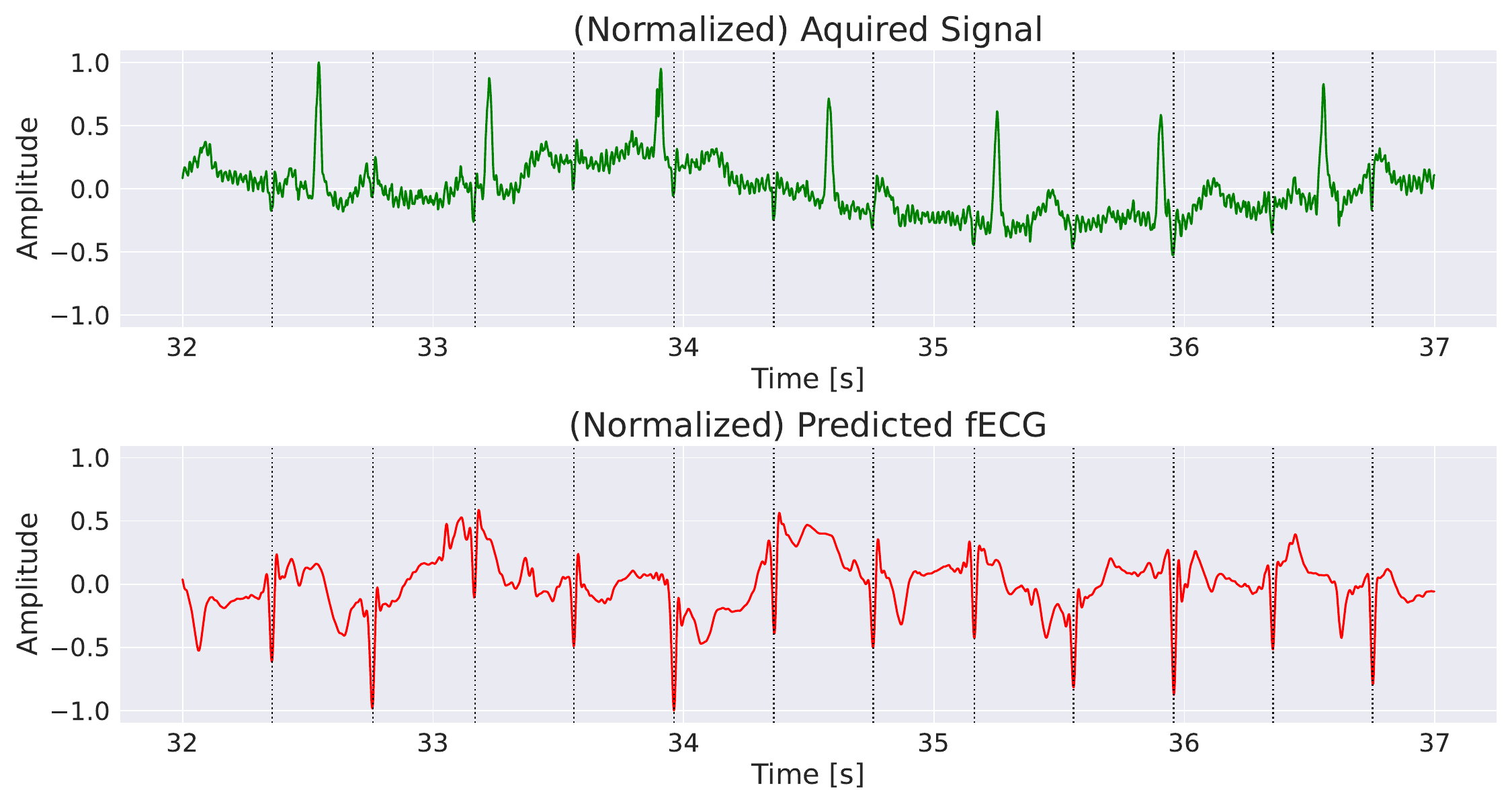}}
    \subfigure[]{\includegraphics[width=\linewidth]{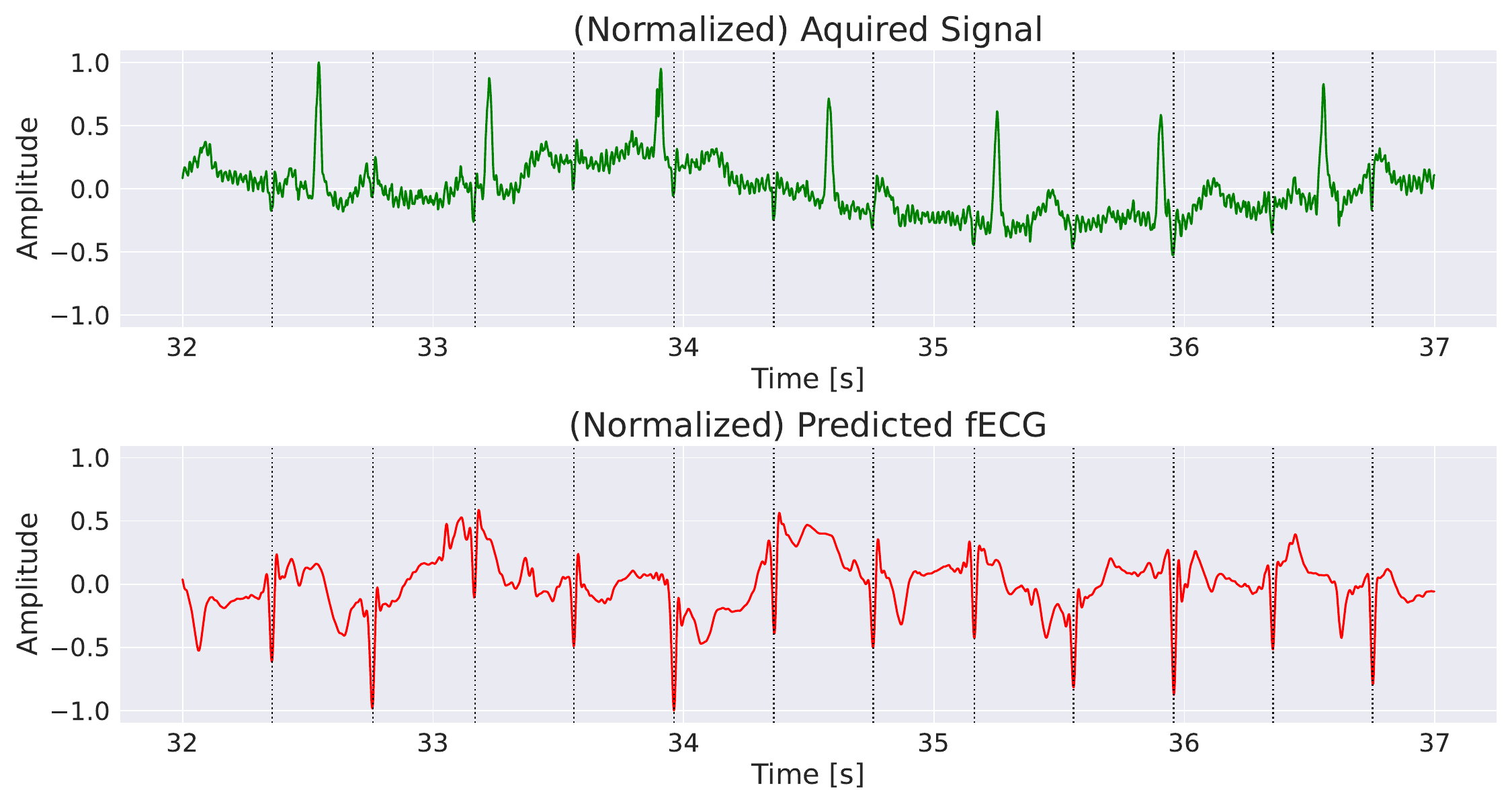}}
    \caption{$\CC$UNet prediction (\textcolor{SeabornRed}{red}) on real data (\textcolor{SeabornGreen}{green}): (a) depicts a signal in which the fECG is visible, while (b) represents a more noisy example. All signals are normalised for a better visualisation. We highlighted the beats with vertical dotted lines.}
     \label{fig:ewam_prediction}
      \vspace{-0.5cm}
\end{figure}

\subsubsection{Dry dataset}
\label{sec:dis-vivo}
\cref{fig:ewam_prediction} illustrates the performance of the proposed $\CC$UNet in extracting the fECG from real data. In both panels (a) and (b), the acquired signals (in green) and the corresponding $\CC$UNet predictions (in red) are shown. Panel (a) presents a case where the fECG is visible in the acquired signal, allowing for a more accurate prediction. In contrast, panel (b) depicts a noisier example, where the extraction process is more challenging. Despite the noise, $\CC$UNet still provides a reasonable estimation of the fECG. 

\section{Discussion}
\label{sec:disc}
The major novelty of this paper lies in proposing a novel deep learning algorithm based on a complex-valued neural network component in order to obtain a reliable and accurate approximation of the fECG signal from single-channel recordings acquired through dry textile electrodes. 

We focused on the development of a network that integrates both amplitude and phase information, denoted as $\CC$UNet. Our findings indicate that this approach not only outperformed the aforementioned networks and traditional methods in the assessed tasks—heart rate determination and fECG extraction across various scenarios, but also exhibited superior generalisation capabilities, as supported by the results from \cref{tab:metriche-physionet}. Furthermore, we established that complex-valued neural networks are more effective in preserving the morphology of the signal compared to their real-valued counterparts.
\subsection{In-silico analysis}
\label{sec:dis-silico}
The \textit{in-silico} dataset purposefully replicates the dry-electrode environment by introducing noise with frequency characteristics matching those observed in surface dry-electrode fECG recordings as demonstrated in \ref{sec:silico}. This allowed the development of a method based on supervised learning. Furthermore, 
 Such a synthetic dataset enables the assessment of both R-peak detection as well as the quality of the fECG extraction, as it allows comparisons against the ground truth. Furthermore, it was beneficial for computing and comparing the evaluation metrics described previously in section \ref{sec:metrics} for the deep learning-based architectures (UNet, AttUNet and $\CC$UNet) and traditional fECG extraction methods (EKF, EKS, and SVD) on the same simulated dataset of 100 subjects not used for the training of the networks. 

As shown in \cref{tab:metriche}, while UNet and AttUNet offer significant improvements over traditional methods, they still exhibit limitations due to their real-valued nature. Both models demonstrate excellent R-peak detection capabilities (F-score and sensitivity values over 98\%), but fECG extraction accuracies are far below that of $\CC$UNet (underlined by the lower PCC values and greater PRD values), showing that these architectures did not fully capture the complexity of the fECG.
$\CC$UNet outperforms all of the other methods in the extraction of an accurate fECG signal, obtaining the lowest PRD (26.2 ± 11.6) and the highest PCC (95.9 ± 5.7). This highlights the importance of retaining both amplitude and phase information. Notably, $\CC$UNet’s PRD is less than half of that of AttUNet, demonstrating a significant reduction in distortion. Additionally, the near-perfect PCC value confirms that the extracted signal maintains a high correlation with the reference signal.

The superior performance of the proposed $\CC$UNet, compared to the more basic UNet and AttUNet, is also evident by visual inspection of \cref{fig:predicted-signals}.
This figure compares the predicted fECGs (red dotted lines) obtained with $\CC$UNet, AttUNet, and UNet, with the ground truth (blue line) to highlight discrepancies in amplitude and morphology. 
Visually, $\CC$UNet (upper plot) exhibits the closest alignment with the original fetal signal, faithfully replicating R-peaks as well as other components such as P- and T-waves. 
In contrast, AttUNet (middle plot) slightly underestimates certain peak amplitudes, while UNet (lower plot) introduces distortions near the QRS complexes. These results underscore the ability of $\CC$UNet to extract the fECG signal very accurately while preserving its morphology and demonstrate once again the importance of complex information used in training for signal reconstruction.

Finally, \cref{tab:snr} and \cref{fig:boxplot} highlight another key strength of the proposed $\CC$UNet: its high resilience to noise, as it maintains strong performances across all SNR levels. In particular, \cref{tab:snr}, which also reports R-peak detection metrics for HR calculation, demonstrates that the extracted fECG maintains sufficiently high quality to ensure excellent detection performance across all SNR levels, including the noisiest cases. As expected, and as shown in \cref{fig:boxplot}, the accuracy of the extracted fECG improves as the SNR increases, as evidenced by a consistent decrease in PRD and a corresponding increase in PCC. 
A practical example of this behavior is shown in  \cref{fig:examples}: when the SNR is high (see panel (a)) - typically ar the late stages of pregnancy when the fetal heart is at its largest- the extracted fECG signal is very accurate and the morphology of the signal is very well preserved. This is confirmed by the high correlation with the reference signal and the low PRD value of 18.6. Conversely, in lower SNR conditions such as those reported in panels (b) and (c)  (e.g.,  $-13.2$ dB and $-18.4$ dB), where the fetal signal is significantly weaker---with the fECG signal in panel (c) not even visible in the acquired data---the estimation becomes less precise. However, despite these challenging conditions, the fundamental components of the fECG remain distinguishable, highlighting the robustness of $\CC$UNet even when the fetal signal is heavily masked by mECG and noise.

\subsection{In-vivo data}
\label{sec:dis-vivo}
\subsubsection{PhysioNet dataset} Our first \textit{in-vivo} evaluation of the proposed model was done using the PhysioNet dataset that contains wet electrode abdominal recordings obtained from pregnant women in clinical settings. Although our model was primarily trained on dry electrode recordings and had no exposure to this dataset, it showcases acceptable performance. This demonstrates $\CC$UNet's capability to generalise in a different setup and data distribution.

The numeric comparison on this \textit{in-vivo} dataset was done regarding R-peak detection, as the dataset contains annotations for the QRS positions. As shown in \cref{tab:metriche-physionet}, $\CC$UNet produces results comparable or even better to the other methods presented in the literature, with F-score of $77.8 \pm 18.6\%$ and sensitivity of $72.1 \pm 22.5\%$, demonstrating its ability to extract the fECG signal. Although EKF achieves a slightly higher F-score ($80.0 \pm 20.0\%$), its higher $\text{HR}_{\text{err}}$ ($15.4 \pm 18.6$ bpm) suggests it misses more fetal R-peaks compared to $\CC$UNet.
UNet's and AttUNet's performances lag behind $\CC$UNet in all three metrics. The substantial drop in F-score and SE for these two networks can be attributed to the shift in data distribution, as the wet electrode recordings exhibit different noise characteristics, to which they were not exposed. Since they use real-value representations, they do not capture phase-related features that are fundamental for fECG extraction, especially in unseen conditions. Furthermore, unlike the \textit{in-silico} dataset in which the noisy environment was well controlled, real recordings can present complex artefacts (electrode motion, biological variability) that are not mitigated by these models, while including the phase by the $\CC$UNet helps to effectively do so.

The PhysioNet dataset does not contain a reference fECG signal, needed to quantify the quality of the fECG extraction. As such, visual inspection is used to ensure a comparison of the obtained fECG using UNet, AttUNet, $\CC$UNet, and benchmark algorithms. \cref{fig:example_Physionet} illustrates an example of the fECG extractions using all the considered methods. It can be seen that UNet and AttUNet fail to generalise well on the \textit{in-vivo} dataset, as their reconstructions appear highly distorted, with missing or misaligned R-peaks. This leads to poor performance in HR approximation metrics. Traditional methods (EKF, EKS, SVD) extract the fECG with different levels and noise suppression. EKF and EKS replicate weak signals, while SVD, which shows clearer QRS complexes, exhibits distortions in the rest of the signal. In comparison, $\CC$UNet preserved the morphology of the fECG better than the other methods. The extracted signal has clear QRS complexes, reduced noise levels, and no missing R-peaks.
The ability to maintain acceptable performance despite strongly varying recording environments (such as amplitude differences, electrode types, and noise characteristics) highlights the robustness of our approach. Unlike traditional methods that usually struggle with domain variation, our model can extract the fECG signal on a completely new dataset, without additional fine-tuning. The superior performance of the proposed model can be associated with its ability to learn complex relationships within data, allowing the thorough separation of fetal, maternal, and noise components, in contrast to traditional methods that rely deeply on assumptions about signal properties, which may not always be true in the real world.

\subsubsection{Dry Dataset} In concluding our analysis, a limited dataset obtained using dry electrodes was employed for visual evaluation. As illustrated in \cref{fig:ewam_prediction}, our proposed method successfully predicts fetal electrocardiograms (fECG) in a real-world context. In the initial example, the fetal beats exhibit minimal amplitude. As such, we decided on highlighting the positions of all the fetal heartbeats using black dotted lines.

The second one depicts a case in which fetal heartbeats are completely masked in mECG and noise, but $\CC$UNet successfully manages to extract its approximation. This is a breakthrough, as real-world dry electrode recordings are already a notoriously challenging scenario. In addition to that, our proposed method relies completely on only one channel, and despite all of these difficulties,  $\CC$UNet maintains good performance, demonstrating high robustness and potential for practical, long-term monitoring solutions. This is a significant achievement, reinforcing the importance of complex data and deep neural networks in further studies regarding fECG extraction.

\subsection{Limitations}
\label{sec:lim}

While the model was primarily trained on an \textit{in-silico} scenario, it also proved its efficiency in \textit{in-vivo} scenarios. This can be attributed to the capability of our model to generalise information, even if the data distribution significantly differs. However, further fine-tuning the model on a more diverse real-world dataset could improve performance and help capture more of the variability of the fECG recordings. Although the proposed method showed high resilience to noise, extremely low SNR values (e.g., first stages of pregnancy when fECG amplitude is really low) could still pose problems. Additionally, while the model was tested on different datasets, a large-scale clinical validation study would be necessary before employing it in clinical settings. Lastly, training the complex-valued UNet requires a higher computational cost than that of training UNet or AttUNet or using traditional algorithms. While it is suitable for a GPU, including it in wearable monitoring devices would need further optimisations and additional studies for obtaining an accurate medical interpretation of the results in real-time.

\section{Conclusion}
\label{sec:conclusion}
In this work, we propose a new method for extracting the fECG signal from recordings of the maternal abdomen. To the best of our knowledge, this is the first work that combines: 
(i) fetal ECG extraction from dry textile electrodes placed on the maternal abdomen, (ii) single-channel recordings of the aforementioned signal, and (iii) complex-valued neural network for isolating the fECG signal. 
We started by analysing several real recordings from non-pregnant women to determine the characteristics of the dry electrode acquisition noise. With this information, we created an \textit{in-silico} dataset to use as training data for our method and compared the results with other well-established methods and architectures. Our proposed model significantly outperforms others across all metrics for the designed scenario.

Moreover, our approach shows better generalisation capabilities than other deep learning methods on both \textit{in-vivo} data and out-of-domain adaptation to wet electrodes.
Compared to traditional methods, our approach better approximates the morphology of a fECG signal, which results in better diagnostics from physicians and healthcare professionals.
Ultimately, our findings might contribute to home-based fetal monitoring, ultimately reducing the number of pregnancy complications.

\section*{References}
\bibliographystyle{IEEEtran}
\bibliography{MyBib}
\makeatletter
\renewcommand{\@bibitem}[1]{\item \Hy@raisedlink{%
  \@highlightscope\@bibitem@hypertarget{cite.#1}%
  \hskip\labelsep\noindent}}
\makeatother

\end{document}